\documentclass[12pt]{article}
 \usepackage{epsfig}
 \def\be{\begin{equation}}
 \def\ee{\end{equation}}
 \def\bea{\begin{eqnarray}}
 \def\eea{\end{eqnarray}}
 \usepackage{graphicx}

 \catcode`\@=11
 \def\lsim{\mathrel{\mathpalette\@versim<}}
 \def\gsim{\mathrel{\mathpalette\@versim>}}
 \def\@versim#1#2{\vcenter{\offinterlineskip
 \ialign{$\m@th#1\hfil##\hfil$\crcr#2\crcr\sim\crcr } }}
 \catcode`\@=12

 \catcode`@=12
\begin{document}
 \thispagestyle{empty}
 \begin{flushright}
 UCRHEP-T601\\
 Apr 2020\
 \end{flushright}
 \vspace{0.6in}
 \begin{center}
 {\LARGE \bf New Insights on Lepton Number\\ and Dark Matter\\}
 \vspace{0.6in}
 {\bf Ernest Ma\\}
 \vspace{0.2in}
{\sl Physics and Astronomy Department,\\ 
University of California, Riverside, California 92521, USA\\}
\end{center}

\begin{abstract}
Dark matter (DM) is usually assumed to be stabilized by a symmetry, which is 
mostly considered to be $Z_2$.  For example, in supersymmetry it is $R$ 
parity, i.e. $(-1)^{3B+L+2j}$.  However, it may be $Z_n$ or $U(1)_D$, and 
derivable from generalized lepton number.  In this context, neutrinos may 
be Majorana or Dirac, and owe their existence to dark matter, i.e. they 
are scotogenic.
\end{abstract}

\section{Dark Matter Prototypes} 
The simplest DM model~\cite{sz85} is to add a real neutral singlet scalar $S$ 
to the Standard Model (SM) with a new $Z_2$ symmetry under which $S$ is odd 
and all other fields are even.  This symmetry is necessary because the 
would-be allowed term $S\Phi^\dagger \Phi$ in the Lagrangian must be 
forbidden, $\Phi$ being the SM Higgs doublet.  It also forbids the possible 
$S \nu_R \nu_R$ term if $\nu_R$ is added for $\nu_L$ to acquire a small 
seesaw Majorana mass.  The next simplest model~\cite{prv08} is to add a 
singlet Majorana fermion $\chi_L$ so that the term $S \bar{\chi}_L \nu_R$ 
is allowed.\\[10pt]

\noindent $^*$ Talk at International Conference on Neutrinos and Dark Matter, Hurghada, Egypt, January 2020.

\newpage
Another DM prototype is for it to generate a radiative Majorana neutrino 
mass, i.e. the scotogenic mechanism.  The simplest one-loop example~\cite{m06} 
adds three Majorana neutral singlet fermions $N_R$ and one scalar doublet 
$\eta = (\eta^+,\eta^0)$ to the SM.  A new $Z_2$ symmetry is again assumed 
under which they are odd and all other fields are even. Hence the tree-level 
terms $\bar{\nu}_L N_R \phi^0$ are forbidden, but $\bar{\nu}_L N_R \eta^0$ 
are allowed.  The same idea also works in some well-known three-loop 
models~\cite{knt03,aks09,gnr13} of neutrino mass.
\begin{figure}[tbh]
\vspace*{-3cm}
\hspace*{-3cm}
\includegraphics[scale=1.0]{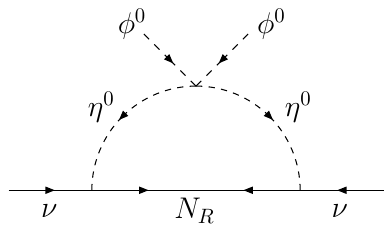}
\vspace*{-21.5cm}
\caption{Radiative seesaw neutrino mass: the scotogenic mechanism.}
\end{figure}

\section{Dark Parity from Lepton Parity}
Even without supersymmetry, the factor $(-1)^{2j}$ may be used to obtain dark 
parity $\pi_D$ from lepton parity $\pi_L = (-1)^L$.  This simple 
observation~\cite{m15} shows that the assignment of lepton parity to new 
particles added to the SM would also determine the dark sector, i.e. no new 
$Z_2$ symmetry is required to obtain exactly the same Lagrangian. 

In the SM, under $\pi_L$, leptons (which are all fermions) are odd and other 
fields are even. In the DM prototypes, $S$ should be assigned odd and $\chi_L$ 
even, so that $S \Phi^\dagger \Phi$ and $S \nu_R \nu_R$ are forbidden, whereas 
$S \bar{\chi}_L \nu_R$ is allowed.  It is clear that the previously imposed 
$Z_2$ dark parity $\pi_D$ is just $(-1)^{2j} \pi_L$.
In the scotogenic model, $N$ and $\eta$ should be assigned even and odd 
repsectively.  Similar assignments are applicable as well in the 
KNT~\cite{knt03}, AKS~\cite{aks09}, and GNR~\cite{gnr13} models.

\section{Lepton Parity with Dark $U(1)_D$}
Instead of assuming lepton parity to begin with, a more general approach is 
to use global $U(1)_L$ and break it softly by two units, but with a particle 
content such that a dark $U(1)_D$ symmetry remains.  Add to the SM three 
pairs of charged fermions $E_L \sim 0, E_R \sim 2$, and two scalar doublets 
$(\eta_1^0,\eta_1^-) \sim 1$, $(\eta_2^{++},\eta_2^+) \sim -1$, plus one 
scalar singlet $\chi^0 \sim -1$, then use the soft term $\bar{E}_L E_R$ to 
break $U(1)_L$ by two units.  A scotogenic Majorana neutrino mass is 
obtained, but $U(1)_D$ remains.  Here $\chi^0$ (mixing slightly with 
$\bar{\eta}^0$) is DM.
\begin{figure}[tbh]
\vspace*{-5cm}
\hspace*{-3cm}
\includegraphics[scale=1.0]{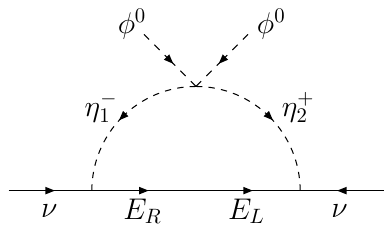}
\vspace*{-21.5cm}
\caption{Scotogenic Majorana neutrino mass with $U(1)_D$.}
\end{figure}

\section{Lepton Number Variants}
The usual theoretical thinking on neutrinos is that they should be Majorana. 
Given that there is still no experimental proof, i.e. no evidence of 
neutrinoless double beta decay, it is time that this idea is re-examined. 
The usual argument goes like this.  For $\nu_L$ to acquire mass, $\nu_R$ 
should be added to the SM, but then $\nu_R$ is allowed to have a large 
Majorana mass, hence $\nu_L$ gets a small seesaw mass and everyone is happy. 
However, $\nu_R$ is a trivial singlet in the SM and its existence is not 
required.

To enforce its existence, the SM should be extended, including gauge $B-L$ 
for example.  In that case, the breaking of $B-L$ by two units would allow 
$\nu_R$ to have a Majorana mass as usual, but breaking it by three units 
would not.  This means that a residual global U(1) remains which protects 
the neutrino as Dirac fermion~\cite{mpr13}.  Depending on the details of 
the new particle content, the new lepton symmetry may be $Z_3$~\cite{mpsz15} 
or $Z_4$~\cite{hr13} or $Z_n (n \ge 5)$.

Combining this recent insight with that on DM, new models of Dirac neutrinos 
and dark matter are possible.  Using gauge $B-L$, instead of having three 
$\nu_R \sim 1$, the theory is also anomaly-free with three right-handed 
neutral singlet fermions transforming as $4,4,-5$~\cite{mp09}.  In that case, 
tree-level Dirac neutrino masses are forbidden, but they may be generated 
radiatively by adding a suitable set of new fermions and scalars.  Three 
recent studies are Refs.~\cite{bccps18,cryz19,jvs19}.

\section{Scotogenic Dirac Neutrino Mass with $Z_n^L$ and $Z_n^D$ ($n \ge 5$)}
To obtain a radiative Dirac neutrino mass induced by dark matter (scotogenic), 
three symmetries are usually assumed~\cite{gs08,fm12}: (A) conventional 
lepton number, where $\nu_{L,R}, N_{L,R}$ have $L=1$, and $\Phi, \eta, \chi$ 
have $L=0$, which is strictly conserved; (B) dark $Z_2$ symmetry, under which 
$N_{L,R}, \eta, \chi$ are odd and others are even, which is strictly 
conserved; and (C) an {\it ad hoc} $Z_2$ symmetry under which $\nu_R, \chi$ 
are odd and all others even, which is softly broken by $\eta^\dagger \Phi 
\chi$.
\begin{figure}[tbh]
\vspace*{-5cm}
\hspace*{-3cm}
\includegraphics[scale=1.0]{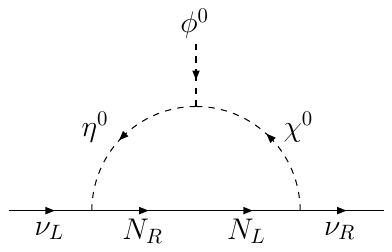}
\vspace*{-21.5cm}
\caption{Scotogenic Dirac neutrino mass.}
\end{figure}

To obtain exactly the same one-loop diagram, it has been shown 
recently~\cite{m19} that a softly broken $U(1)_L$ by itself will do the 
job. Consider the following particle content, as shown in Table 1.
\begin{table}[tbh]
\centering
\begin{tabular}{|c|c|c|c||c|c|c|}
\hline
fermion/scalar & $SU(2)$ & $U(1)_Y$ & $U(1)_L$ & {\bf *} & $Z_n^L$ & 
$Z_n^D$ \\ 
\hline
$(\nu,e)_L$ & 2 & $-1/2$ & 1 & 1 & $\omega$ & 1 \\ 
$e_R$ & 1 & $-1$ & 1 & 1 & $\omega$ & 1 \\ 
$\nu_R$ & 1 & 0 & $x$ & $-n+1$ & $\omega$ & 1 \\ 
$N_L$ & 1 & 0 & $y$ & $2-n$ & $\omega^2$ & $\omega$ \\ 
$n_R$ & 1 & 0 & $y$ & $2-n$ & $\omega^2$ & $\omega$ \\ 
\hline
$\Phi = (\phi^+,\phi^0)$ & 2 & 1/2 & 0 & 0 & 1 & 1 \\ 
$\eta = (\eta^+,\eta^0)$ & 2 & 1/2 & $y-1$ & $1-n$ & $\omega$ & $\omega$ \\ 
$\chi^0$ & 1 & 0 & $y-x$ & 1 & $\omega$ & $\omega$ \\ 
\hline
\end{tabular}
\caption{Fermion and scalar content for scotogenic Dirac neutrino mass.}
\end{table}

Here $x \neq 1$ is imposed so that $\nu_R$ does not couple to $\nu_L$ at tree 
level.  To connect them in one loop, the trilinear 
$\bar{\eta}^0 \phi^0 \chi^0$ term must break $U(1)_L$ softly by $x-1$. The 
$y$ charge of $N_{L,R}$ must not be $\pm 1$ or $\pm x$ to avoid 
undesirable couplings to $\nu_{L,R}$.  The soft terms $N_L N_L$ or $N_R N_R$ 
would break $U(1)_L$ by $2y$, $\nu_R \nu_R$ by $2x$, $N_R \nu_R$ by $x+y$, 
$\bar{N}_L \nu_R$ by $x-y$, and $\chi^0 \chi^0$ by $2(y-x)$.  They should be 
absent, hence they must not be zero or divisible by $x-1$.  The column 
denoted by {\bf *} shows a class of solutions where $U(1)_L$ breaks to $Z_n$, 
i.e. $x=-n+1$ and $y=2-n$.

If $n=3$, then $x+y=-3$.  If $n=4$, then $2y=-4$.  Hence $n=3,4$ are ruled 
out.  Any $n \ge 5$ works.  This results in two related symmetries: (I) 
$Z_n^L$ lepton symmetry under which $\nu_{L,R}, e_{L,R}, \eta, \chi \sim 
\omega$ and $N_{L,R} \sim \omega^2$, where $\omega^n = 1$; (II) $Z_n^D$ 
dark symmetry, derivable from $Z_n^L$ by multiplying it by $\omega^{-2j}$ 
where $j$ is the particle's spin.  As a result, $\nu_{L,R}, e_{L,R} \sim 1$ 
and $N_{L,R}, \eta, \chi \sim \omega$.  This is the Dirac generalization 
of $\pi_D = (-1)^{2j} \pi_L$ for Majorana neutrinos.

In a renormalizable theory, the $Z_n$ symmetry is not simply realizable. 
For $n \ge 5$, $(\chi^0)^n$ is not admissable. Hence the Lagrangian 
actually has a redefined $U(1)_L$ symmetry under which 
$\nu_{L,R}, e_{L,R}, \eta, \chi \sim 1$ and $N_{L,R} \sim 2$. The dark 
symmetry is then $U(1)_D$ where it is derived from $U(1)_L$ by 
subtracting $2j$, i.e. $\nu_{L,R},e_{L,R} \sim 0$ and $N_{L,R}, \eta, 
\chi \sim1$.

If $Z_n$ symmetry is desired, the scalar sector must be expanded.  If $n=5$, 
let $\sigma \sim 3$ and $\kappa \sim 7$ be added.  Then the terms 
$\chi^3 \sigma^*, \chi^2 \sigma, \chi \sigma^2 \kappa^*$, and 
$\kappa N_R \nu_R$ are allowed.  Together they would enforce $Z_5^L$ and 
$Z_5^D$.

A possible variation is to add $\zeta \sim n$ and require $U(1)_L$ to be 
spontaneously broken in the $\zeta^* \eta^\dagger \Phi \chi$ term, thereby 
yielding a massless Goldstone boson, i.e. the diracon~\cite{bv16}, as the 
analog of the majoron for Majorana neutrinos.  A further application is 
to allow $\zeta$ to couple anomalously to exotic color fermion triplets or 
a color fermion octet~\cite{dm00}.  The diracon becomes the QCD axion and 
$U(1)_L$ is extended Peccei-Quinn symmetry, as proposed long ago~\cite{s88} 
for Majorana neutrinos, and very recently for Dirac 
neutrinos~\cite{prsv19,b19}.

\section{Scotogenic Dirac Neutrino Mass with $Z_3^D$}
The $N_{L,R}$ fermion singlets may be replaced by $(E^0,E^-)_{L,R}$ fermion 
doublets, as shown in Table 2.
\begin{table}[tbh]
\centering
\begin{tabular}{|c|c|c|c||c|c|c|}
\hline
fermion/scalar & $SU(2)$ & $U(1)_Y$ & $U(1)_L$ & {\bf **} & $L$ & 
$Z_3^D$ \\ 
\hline
$(\nu,e)_L$ & 2 & $-1/2$ & 1 & 1 & $1$ & 1 \\ 
$e_R$ & 1 & $-1$ & 1 & 1 & $1$ & 1 \\ 
$\nu_R$ & 1 & 0 & $x$ & $-2$ & $1$ & 1 \\ 
$(E^0,E^-)_L$ & 2 & $-1/2$ & $y$ & $2$ & $1$ & $\omega$ \\ 
$(E^0,E^-)_R$ & 2 & $-1/2$ & $y$ & $2$ & $1$ & $\omega$ \\ 
\hline
$\Phi = (\phi^+,\phi^0)$ & 2 & 1/2 & 0 & 0 & 0 & 1 \\ 
$\eta = (\eta^+,\eta^0)$ & 2 & 1/2 & $x-y$ & $-4$ & $0$ & $\omega^{-1}$ \\ 
$\chi^0$ & 1 & 0 & $y-1$ & 1 & $0$ & $\omega$ \\ 
\hline
\end{tabular}
\caption{Fermion and scalar content for scotogenic Dirac neutrino mass with 
$Z_3^D$ dark symmetry.}
\end{table}

This construction eliminates the existence of many fermion bilinears except 
$\nu_R \nu_R$ and $\bar{\nu}_L E_R^0 + e^+_L E^-_R$.  Hence only $2x$ and 
$y-1$ must not be zero or divisible by $x-1$.  Also $y \neq x$ is required. 
Now $Z_3^D$ is possible as shown in the column denoted by {\bf **}. Here 
$U(1)_L$ is broken by the soft $\Phi^\dagger \eta \chi$ and $\chi^3$ terms. 
However the dimension-four term $\chi^0 \nu_R \nu_R$ is not allowed by the 
original $U(1)_L$ even though it is allowed by $Z_3^D$.  Hence the usual 
lepton assignment holds: $L=1$ for $\nu_{L,R}, E_{L,R}$ and $L=0$ for all 
scalars. [Note that in $Z_3$, if $\nu_R \sim \omega$, then $\chi \sim \omega$ 
or $\omega^2$.  Hence either $\chi \nu_R \nu_R$ or $\chi^* \nu_R \nu_R$ must 
exist and $\chi$ cannot be stable.]

In this example, $U(1)_L$ is anomalous.  To make it anomaly-free, the three 
copies of $\nu_R$ with charge $-2$ should be changed to 1.  The difference 
is then $3[1-(-2)]=3[3]=9$ for the sum of the charges, and 
$3[1-(-8)]=3[9]=27$ for the sum of the cubes of the charges.  This may be 
accomplished with singlet right-handed fermions $\psi_{2,3,4}$ with 
charges $-2,3,-4$.  For 9 copies of $\psi_3$ and 3 copies of $\psi_{2,4}$, 
$[3[3(3)+(-2)+(-4)]=3[3]=9$ and $3[3(27)+(-8)+(-64)]=3[9]=27$. 
Add $\zeta_{3,6}$ to break $U(1)_L$ to $Z_3^D$.  Then $\psi_{2,3,4}$ have 
$L=1,0,-1$.

\begin{figure}[tbh]
\vspace*{-5cm}
\hspace*{-3cm}
\includegraphics[scale=1.0]{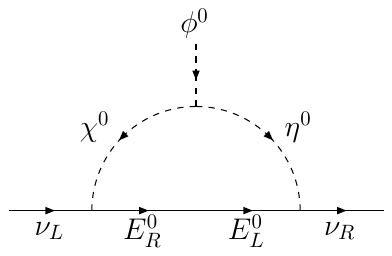}
\vspace*{-21.5cm}
\caption{Scotogenic Dirac neutrino mass with $Z_3^D$.}
\end{figure}

\section{Concluding Remarks}
The notion of generalized $U(1)_L$ is useful in connecting leptons to dark 
matter. If it breaks softly to $Z_2$, then many DM prototype models may be 
understood in terms of lepton parity $\pi_L$ (conserved for Majorana 
neutrinos) alone, with dark parity $\pi_D = (-1)^{2j} \pi_L$. 

In scotogenic models, $\pi_L$ and $U(1)_D$ are possible together.  For Dirac 
neutrinos, softly broken $U(1)_L$ may also lead to $Z_n^L$ and $Z_n^D$ with 
$n \ge 5$, or redefined $U(1)_L$ and $U(1)_D$.  An example of $Z_3^D$ and 
conventional $L$ is also possible.

\section{Acknowledgement}
I thank Shaaban Khalil and all other local organizers for their great effort 
and hospitality regarding NDM-2020.  This work was supported in part by the 
U. S. Department of Energy Grant No. DE-SC0008541.

\bibliographystyle{unsrt}

\end{document}